\documentclass{emulateapj}
\usepackage{apjfonts,graphics}
\shorttitle{A1689 and Triaxial Dark Halos}
\shortauthors{Oguri et al.}
\begin{document}
\title{
Can the Steep Mass Profile of A1689 Be Explained by a Triaxial Dark Halo?$^1$}
%
\author{
Masamune Oguri,\altaffilmark{2,3}
Masahiro Takada,\altaffilmark{4} 
Keiichi Umetsu,\altaffilmark{5}
and Tom Broadhurst\altaffilmark{6}
}
\altaffiltext{1}{Based in part on data collected at the Subaru Telescope,
  which is operated by the National Astronomical Society of Japan}
\altaffiltext{2}{Princeton University Observatory, Peyton Hall,
Princeton, NJ 08544}
\altaffiltext{3}{Department of Physics, University of Tokyo, Hongo
7-3-1, Bunkyo-ku, Tokyo 113-0033, Japan}
\altaffiltext{4}{Astronomical Institute, Tohoku University, Sendai
980-8578, Japan} 
\altaffiltext{5}{Institute of Astronomy and Astrophysics, Academia
Sinica,  P.~O. Box 23-141, Taipei 106,  Taiwan, Republic of China}
\altaffiltext{6}{School of Physics and Astronomy, Tel Aviv University,
Israel} 
%
%
\begin{abstract}
The steep mass profile of A1689 derived from recent detailed lensing
observations is not readily reconciled with the low concentration halos
predicted by the standard cold dark matter (CDM) model. However, halo
triaxiality may act to bias the profile constraints derived assuming a
spherically symmetric mass distribution, since lensing relates only to
the projected mass distribution.  A degree of halo triaxiality is
inherent to the CDM structure formation, arising from the collision-less
nature of the dark matter.  Here we compare the CDM-based model
predictions of triaxial halo with the precise lensing measurements of
A1689 based on the Advanced Camera for Surveys/Hubble Space Telescope
and Subaru data, over a wide range of $10{\rm kpc}\lesssim r \lesssim
2{\rm Mpc}$.  The model lensing profiles cover the intrinsic spread of
halo mass and shape (concentration and triaxiality), and are projected
over all inclinations when comparing with the data.  We show that the
model parameters are only weakly constrained and strongly degenerate
mainly because of the lack of information along the line of sight.  In
particular, the limits on the concentration parameter become less
restrictive with increasing triaxiality.  Yet, by comparing the obtained
constraints with expected probability distributions for the axis ratio
and concentration parameters computed from numerical simulations, we
find that $\sim 6\%$ of cluster-size halos in the CDM model can match
the A1689 lensing observations at the 2-$\sigma$ level, corresponding to
cases where the major-axis of the halo is 
aligned with the line
of sight. Thus halo triaxiality could reduce the apparent discrepancy
between theory and observation.  This hypothesis needs to be further
explored by a statistical lensing study for other clusters as well as by
complementary three-dimensional information derived using $X$-ray,
kinematics, and Sunyaev-Zel'dovich effect observations.
\end{abstract}
 
 
\section{Introduction}

The current standard model of structure formation, in which the universe
is dominated by Cold Dark Matter (CDM), predicts that dark matter halos
have an inner cusp that is shallower than a singular isothermal sphere
model. Specifically, the radial density profile can be well fitted by an
NFW profile $\rho(r)\propto r^{-1}(1+r/r_s)^{-2}$ \citep{navarro97}. An
important parameter of this model is the concentration parameter $c_{\rm
vir}$,  the ratio of the virial radius to the scale radius
$r_s$. For massive clusters, it is predicted to be $\sim 4$
\citep[e.g.,][]{bullock01}. These are quite important predictions that
should be confronted with observations.  

Now it is becoming possible to {\it directly} test the NFW predictions
using gravitational lensing. A poster child example for this is A1689:
The inner mass distribution is tightly constrained from more than 100
multiple images of background galaxies discovered from the spectacular
deep Advanced Camera for Surveys (ACS)/Hubble Space Telescope (HST) data
\citep[][ hereafter B05b]{broadhurst05b}, while the larger scale mass
distribution up to the virial radius is obtained from the weak lensing
measurements based on the wide-field Suprime-Cam/Subaru data 
\citep[][ hereafter B05a]{broadhurst05a}. The two-dimensional mass
profile of A1689, reconstructed from the combined lensing information,  
does continuously flatten towards the center like an NFW profile, but
the fitting to the NFW predictions leads to a surprisingly high
concentration $c_{\rm vir}=14\pm1.5$ (B05a), compared with the expected
value $c_{\rm vir}\sim 4$. Such a high concentration is also seen in
other clusters, e.g., MS2137-23 \citep{gavazzi03} and possibly CL
0024+1654 \citep{kneib03}. These results may represent a new crisis in
the current standard CDM model, offering a very rewarding issue to
further explore.  

However, the constraints above are obtained by deprojecting the
two-dimensional lensing information under assumption of the spherically
symmetric mass distribution. What the CDM model does predict is that
dark halos are quite triaxial rather than spherical as a natural
consequence of the collision-less dark matter nature and the filamentary
nature of structure formation (\citealt[][ hereafter JS02]{jing02};
\citealt{lee05}). Therefore the results obtained assuming spherical
symmetry might be biased \citep[e.g., see][ for the similar discussion
on A1689]{miralda95}. In fact, \citet{clowe04} argued that the halo
triaxiality affects both mass and concentration parameter measurements
from gravitational lensing using $N$-body simulation results of
clusters. \citet{gavazzi05} also investigated the importance of the halo
triaxiality and argued that a high concentration of MS2137-23 could be
explained by a halo having the major axis oriented toward the
line-of-sight.  

In this paper, we study the mass distribution of A1689 based on triaxial
dark halo model. Specifically, we adopt the triaxial halo model of JS02,
and focus on whether the apparent steep mass profile of A1689 can be
ascribed to triaxiality of the mass distribution. Throughout this
paper we assume a concordance cosmology with the matter density
$\Omega_M=0.3$, the cosmological constant $\Omega_\Lambda=0.7$, the
dimensionless Hubble constant $h=0.7$, and the normalization of matter
power spectrum $\sigma_8=0.9$. Note that one arcminute corresponds to
the physical scale of $129h^{-1}$kpc for A1689 (redshift $z=0.18$). 

\section{A simple estimation of the halo triaxiality effect on lensing
 measurements} 
\label{sec:test}

Before going to the analysis of A1689, we make a simple test to
demonstrate how important the halo triaxiality is in constraining mass
profiles from a two-dimensional lensing measurement. 
The analysis is somewhat similar to that done by
\citet{clowe04} who used high-resolution $N$-body simulations of massive
clusters. Here we instead use an analytic model of aspherical dark halos.

\begin{deluxetable*}{cccccc}
\tablewidth{0pt}
\tablecaption{Best-fit parameters\label{table:fit} for the model fitting
 to A1689}
\tablehead{\colhead{$\chi^2$} & \colhead{$R_e[{\rm Mpc}]$} &
 \colhead{$c_e$} & \colhead{$a/c$} & \colhead{$b/c$} & 
\colhead{$M_{\rm vir}[10^{15}M_\odot]$}}
\startdata
$\chi^2_{\rm obs}$ & $1.80^{+0.08}_{-0.55}$ & $7.6^{+1.0}_{-5.8}$ 
& $0.1(<0.75)$ & $0.85(>0.32)$ & $1.8^{+0.4}_{-0.7}$
\\
$\chi^2_{\rm obs}+\chi^2_{\rm ar}$ & $1.50^{+0.11}_{-0.12}$ &
 $5.2^{+1.8}_{-1.9}$ & $0.45^{+0.07}_{-0.10}$ &
 $0.6^{+0.20}_{-0.15}$ & $1.7^{+0.3}_{-0.2}$
\enddata
\tablecomments{Best-fit parameters and the 1-$\sigma$ errors are
 presented, for the triaxial halo model fitting to the A1689 lensing
 information. Note that the virial mass $M_{\rm vir}$ is not a model
 parameter, but is derived from the constrained model parameters $R_e$,
 $c_e$, $a/c$, and $b/c$ (see text for details). The lower row
 shows the results when the theoretically expected probability
 distributions for the axis ratios are employed as the priors of the
 fitting. For some parameters, only a lower or upper limit is obtained.}
\end{deluxetable*}

\begin{figure}
\epsscale{1.1}
\plotone{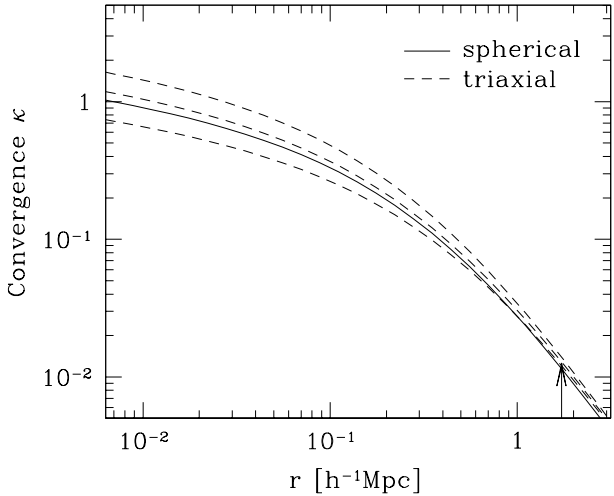}
\caption{Convergence profiles of the triaxial halo with the virial mass 
 $M_{\rm vir}=10^{15}h^{-1}M_\odot$, the concentration parameter
 $c_e=1.15$, and the triaxial axis ratios of $a/c=0.4$ and $b/c=0.7$. 
 The halo is placed at $z_l=0.3$, and we assume the source redshift of 
 $z_s=1.0$. We consider the projection along each of the three principal
 axes: from upper to lower the dashed lines show profiles projected
 along $z$, $y$, and $x$ (see  eq. [\ref{gnfw}]), respectively. 
 The convergence profile of the corresponding spherical NFW halo is also
 plotted by the solid line for comparison (see text for details). The
 vertical arrow indicates the virial radius.  
\label{fig:kap}}
\end{figure}
We consider a triaxial halo with the virial mass $M_{\rm
vir}=10^{15}h^{-1}M_\odot$\footnote{The virial mass is defined 
by spherically averaging the halo mass distribution (the triaxial mass
profile for our case) around the halo center and then by finding the
sphere inside which the mean overdensity reaches $\Delta_{\rm vir}$
predicted in the top-hat spherical collapse model.},  placed at
$z_l=0.3$, and adopt the model mass profile given in JS02:  
\begin{equation}
 \rho(R)=\frac{\delta_{\rm ce}\rho_{\rm crit}(z)}
{(R/R_0)(1+R/R_0)^{2}},
\label{gnfw}
\end{equation}
\begin{equation}
 R^2\equiv c^2\left(\frac{x^2}{a^2}+\frac{y^2}{b^2}
+\frac{z^2}{c^2}\right)\;\;\;(a\leq b\leq c).
\label{rdef}
\end{equation}
We adopt typical model parameters for a halo of $10^{15}h^{-1}M_\odot$:
the triaxial axis ratios are $a/c=0.4$ and $b/c=0.7$, and the
concentration parameter $c_e\equiv R_e/R_0$, where $R_e$ is defined such
that the mean density enclosed within the ellipsoid of the major axis
radius $R_e$ is $\Delta_e\Omega(z)\rho_{\rm crit}(z)$ with $\Delta_e =
5\Delta_{\rm vir}\left(c^2/ab\right)^{0.75}$, is chosen to be
$c_e=1.15$. We have checked that the spherically-averaged radial mass
profile of the triaxial halo is quite similar to the spherical NFW
profile that is specified by the virial radius $r_{\rm vir}=R_e/0.45$,
as proposed in JS02 (see Figure 14 in JS02), and the concentration
parameter $c_{\rm vir}=4$\footnote{Since the overdensity in the triaxial
model, $\Delta_{\rm e}$, is at least 5 times larger than the spherical
overdensity $\Delta_{\rm vir}$, the concentration parameter in the
triaxial model tends to be smaller than those in the spherical
model (approximately given as $c_e\approx 0.45c_{\rm vir}$). See JS02 for
details.}.  However, it is non-trivial for these 
triaxial and spherical models whether to yield similar lensing maps as a
result of the line-of-sight
projection\footnote{
The lensing convergence field
$\kappa(\mathbf{r})$ is given in terms of the surface mass density
$\Sigma(\mathbf{r})$ as $\kappa(\mathbf{r})\equiv
\Sigma(\mathbf{r})/\Sigma_{\rm cr}$, where $\Sigma_{\rm cr}$ is the
lensing critical density specified for a background cosmology and lens
and source redshifts \citep[see][]{schneider92}.}.  To make this clear,
Figure \ref{fig:kap} compares the circularly-averaged convergence
profiles for the spherical and triaxial halos.  For the triaxial halo,
we consider the projection along each of the three principal axes. It is
clear that the surface mass density of the triaxial halo depends
strongly on the projection direction.  Therefore it is quite likely that
adopting a spherical halo model causes a bias in estimating the mass and
profile parameters for an individual cluster in reality.

\begin{figure}
\epsscale{1.1}
\plotone{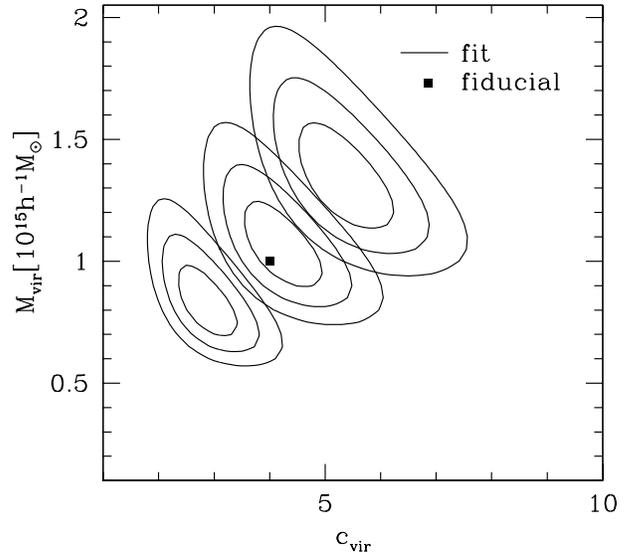}
\caption{Constraint contours in the virial mass and halo concentration 
 parameter space, obtained by fitting the mock data of triaxial halos to
 the spherical NFW halo model. The contours show 68\%, 95\%, 99.7\%
 confidence limits (corresponding to $\Delta\chi^2=2.3, 6.17 $  and
 $11.8$, respectively). From left to right, the constraint contours from the
 convergence profiles of the triaxial halo projected along the principal
 axes $x$, $y$, and $z$ (as in Figure \ref{fig:kap}), respectively, are
 shown. For comparison, the square symbol shows the best-fitting model for the
 convergence profile obtained by projecting the {\em a priori}
 spherically-averaged mass profile of the input triaxial halo. 
\label{fig:chi}} 
\end{figure}
To see this more clearly, we perform the following test. First, we
generate  an ``observed'' surface mass density profile: We consider 20
bins logarithmically spacing over the range $r=[10^{-2},1]h^{-1}{\rm
Mpc}$, and generate the convergence profile
$\kappa(r)$, where the mean value for each bin is taken from the
triaxial halo model and the Gaussian random error of standard deviation
$\Delta(\log_{10}\kappa)=0.1$ is added to each bin. Then, assuming the
spherical NFW density profile, we constrain the virial mass ($M_{\rm
vir}$) and halo concentration  parameter ($c_{\rm vir}$) by fitting the
model predictions to  the ``observed'' profile. The constraint contours
in the $M_{\rm vir}-c_{\rm vir}$ plane are shown in Figure
\ref{fig:chi}, demonstrating that the best-fit parameters depend
strongly on the projection direction. For example, the convergence
profile projected along the major (minor) axis yields a significant
overestimation (underestimation) by $20-30\%$ in {\it both} the mass and
concentration parameters. It should be noted that the bias direction is
orthogonal to the degeneracy direction of the error ellipse, implying
the systematics is very important.  In fact, we fail to recover the
model parameters of the spherically-averaged triaxial profile at more
than 3-$\sigma$ level when the halo is projected along the major- or
minor-axis directions.  

\section{Application to A1689}
\label{sec:mp}

In this section, we indeed make a quantitative interpretation of the
lensing measurements of A1689 in terms of the triaxial halo model in
order to derive more reliable constraints on the three-dimensional mass
distribution than does assuming the spherical halo model. Our particular
attention is paid to whether or not the steep mass profile claimed in
B05a  can be reconciled with the CDM-based triaxial halo model of JS02.  

Following the method of \citet{oguri03}, the model lensing convergence
field for a triaxial halo is specified by 7 parameters. As explained
around equation (\ref{gnfw}), the {\it virial} radius of the triaxial
halo modeling $R_e$, the concentration parameter $c_e$, and the axis
ratio parameters $a/c$ and $b/c$ are considered. Further, we adopt three
angle parameters to specify the halo orientation: $\theta$, $\phi$
\citep[see Figure 1 in][]{oguri03} and the third angle between major
axis of an ellipse of the projected mass distribution and the RA
direction on the sky.  Note that the parameter $\theta$ gives the angle
between the major axis of the triaxial halo and the line-of-sight
direction.  We again note that, in this model, the virial mass $M_{\rm
vir}$  is not a free parameter but is derivable from the model parameters
($R_e$, $c_e$, $a/c$, $b/c$). Throughout this paper, we assume $\langle
z_s\rangle=1$ for the mean redshift of source galaxies as done in B05a.  

We then constrain the 7 model parameters from $\chi^2$ fitting to
the observed convergence map obtained from the ACS/HST and 
Suprime-Cam/Subaru data (e.g. see B05a for the similar method):
\begin{equation}
 \chi^2_{\rm obs}=\chi^2_{\rm HST}+\chi^2_{\rm Subaru},
\label{eqn:chi2}
\end{equation}
where we adopt flat priors of $0.1\le a/c(\le 1)$ and $0.1\le b/c(\le
1)$ for the axis ratios because a small axis ratio such as $a/c\le 0.1$
is unrealistic for a virialized halo.  For the ACS data, we use the
circularly-averaged profile of the obtained convergence map that is
given in 12 bins linearly spacing over the range
$r=[0\farcm038,0\farcm97]$ (see Figure 22 in B05b or Figure 3 in
B05a)\footnote{It is not straightforward to use the two-dimensional mass
map obtained from the ACS strong lensing analysis because of the complex,
non-linear error propagation. The bin width of the one-dimensional
convergence profile we use here is sufficiently broad to ensure that the
errors between different bins are independent (B05b).}. Hence, the 
$\chi^2_{\rm HST}$ is given by
\begin{equation}
 \chi^2_{\rm HST}=\sum_{i=1}^{12}
\frac{\left(\bar{\kappa}_i^{\rm m}-\kappa_i\right)^2}{\sigma_i^2}, 
\end{equation}
where $\bar{\kappa}^{\rm m}_i$ is the model prediction of the triaxial
halo for the $i$-th radial bin, computed as in Figure \ref{fig:kap}, 
and the $\kappa_i$ and $\sigma_i$ are the observed value and
$1$-$\sigma$ error, respectively. As for the Subaru data, we use the
two-dimensional convergence map $\kappa(\mathbf{r})$ reconstructed from
the joint measurements of the weak lensing distortion and magnification
bias effects on background galaxies (Umetsu et al. 2005, in
preparation).  The map is given on $21\times 17(=357)$ grids for the
area of $\approx 30'\times 24'$. Note that the wide-field Subaru data
allows us to probe the mass distribution on larger scales $\gtrsim 1'$
covering up to the virial radius $\sim 20'$ (B05a), thus ensuring that
the ACS and Subaru information are independent. The pixel-pixel
covariance matrix defined as $V_{ij}\equiv
\langle\delta\kappa(\mathbf{r}_i)~  \delta\kappa(\mathbf{r}_j)\rangle$
can be accurately estimated based on the maximum likelihood map-making
method, assuming that the error of weak lensing distortion arises 
from the random intrinsic ellipticities of background galaxies, while
the magnification bias error is the Poissonian noise of the number
counts (\citealt{schneider00}; see Umetsu et al. in preparation for the
details). The $\chi^2_{\rm Subaru}$ can be therefore expressed as 
\begin{equation}
 \chi^2_{\rm Subaru}=\sum_{i,j} \left[\kappa^{\rm m}(\mathbf{r}_i)-
\kappa(\mathbf{r}_i)\right]
(V^{-1})_{ij}\left[\kappa^{\rm m}(\mathbf{r}_j)-\kappa(\mathbf{r}_j)\right],
\end{equation}
where $\mathbf{V}^{-1}$ is the inverse of the covariance matrix.  

Table \ref{table:fit} lists the best-fit values for the 4 model
parameters of our interest as well as the virial mass derived, with the
1-$\sigma$ error including the other 6 parameter uncertainties
(estimated from $\Delta\chi^2\equiv \chi^2-\chi^2_{\rm min}=1$ in the 7
parameter space). The minimum $\chi^2$ is $\chi^2_{\rm min}/{\rm
dof}=378/362$ for 362 degrees of freedom, thus showing an acceptable
fit. It is clear that the halo triaxiality significantly weakens
constraints on the halo concentration parameter $c_e=7.6^{+1.0}_{-5.8}$,
compared to the constraint $c_{\rm vir}=14\pm1.5$  ($c_e\approx
0.45c_{\rm vir}$) derived when {\em a priori} assuming the
spherical NFW model (B05a). In fact, a smaller concentration compatible
with the theoretical expectation $c_e=1-2$ is marginally allowed at 
1-$\sigma$ level.  

\begin{figure*}[t]
\epsscale{1.1} \plotone{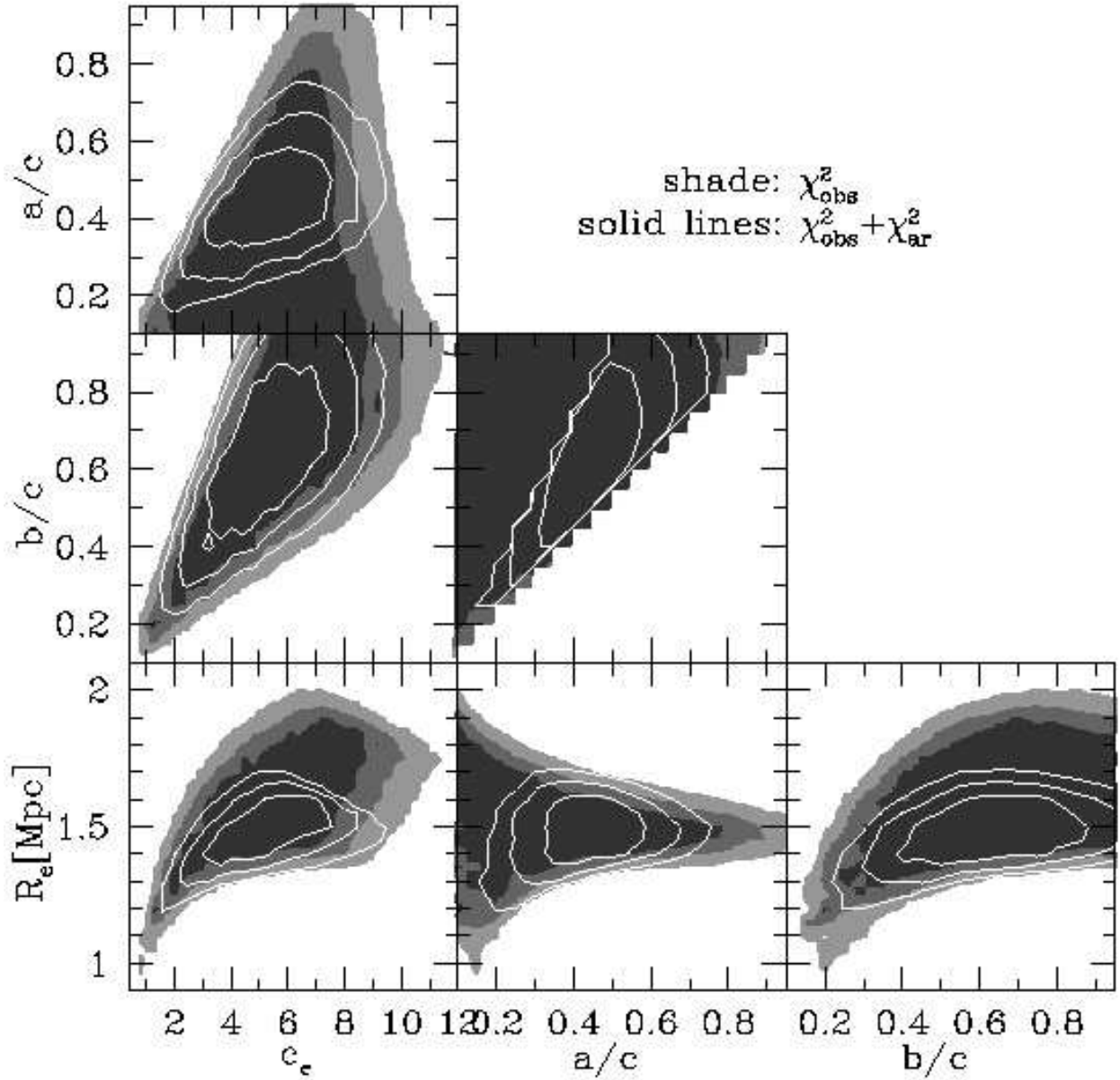} \caption{Projected constraint
contours in the two parameter subspace from the 7 parameter space for
the triaxial halo model, matching the two-dimensional mass distribution
of A1689 reconstructed from the ACS/HST and Subaru data.
The dark shaded regions show the $68\%$, $95\%$ and $99.7\%$ confidence
intervals corresponding to $\Delta\chi^2=2.3$, $6.17$, and $11.8$ in the
7 parameter space, respectively.  It is clear that the parameter
 constraints are significantly weakened by inclusion of the halo
 triaxiality and show strong degeneracies in each parameter space. 
Note that the concentration parameter of the triaxial halo model, $c_e$, 
is approximately related to that of the spherical model via $c_e\simeq
0.45 c_{\rm vir}$ (JS02). The solid contours show the improved
 constraints when the CDM-based theoretical predictions of probability
 distributions for the axis ratios $a/c$ and $b/c$ are added as the
 prior to the $\chi^2$-fitting (see text for details). However, a broad
 range of the parameters are still allowed mainly because of a lack of
 the halo shape information along the line-of-sight.
\label{fig:chi_tri}}
\end{figure*}

\begin{figure}[t]
\epsscale{1.} \plotone{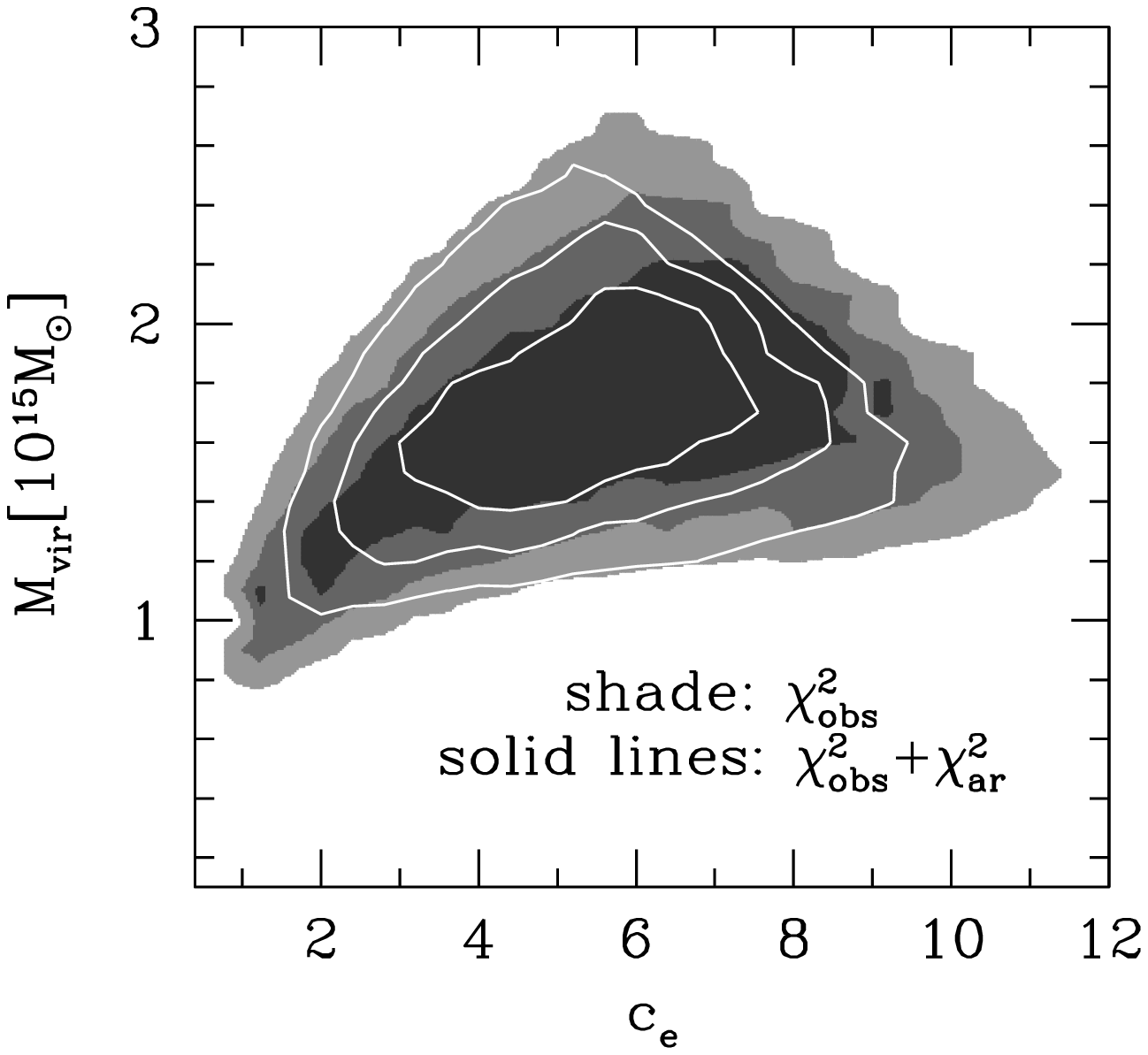} \caption{As in the previous figure, 
the constraint contours are shown in the $c_e-M_{\rm vir}$ plane for the
 triaxial halo model, where the virial mass $M_{\rm vir}$ is derived
 from the model parameters constrained in Figure \ref{fig:chi_tri}.
 Note that the concentration parameter of the triaxial halo model,
 $c_e$, is approximately related to that of the spherical model via
 $c_e\simeq 0.45 c_{\rm vir}$ (JS02).  
 \label{fig:chi_mc}}
\end{figure}

This is more clearly explored in Figure \ref{fig:chi_tri}, which shows
the error contours in the two parameter subspace. We also show the error
contours in the $c_e-M_{\rm vir}$ plane in Figure \ref{fig:chi_mc} to
compare our result with Figure 3 of B05a. Apparently, the parameter
constraints are significantly degenerate. For example, as a triaxial
halo with smaller $a/c$ is considered, a broader range of the halo
concentration is allowed, implying that the lensing strength is as
quite sensitive to the halo triaxiality as is to the halo concentration
$c_e$.  However, the axis ratios can be hardly constrained, obviously
because of a lack of the halo shape information along the line-of-sight: 
Any values of $a/c$ and $b/c$ are allowed to within 3-$\sigma$ level,
though profiles close to spherical models ($a/c\sim 1$) are relatively
disfavored because the Subaru convergence map is not circularly
symmetric. More intriguing is that the degeneracy direction in the
$c_e-M_{\rm vir}$ plane (Figure \ref{fig:chi_mc}) is almost orthogonal
to that for the spherical case (Figure 3 in B05a), as illustrated in
Figure \ref{fig:chi}. As a result, the mass constraint can be reconciled
with the mass estimate derived from the $X$-ray observation
\citep[$\approx 1.0\times 10^{15}M_\odot$ in][]{andersson04} at
3-$\sigma$ level. Thus the halo triaxiality could resolve the mass
discrepancy reported in the literature (\citealt{miralda95,andersson04};
B05b), as we will again discuss later in more detail.   

To improve the weak constraints above, we employ the prior knowledge of
what kinds of halo shapes are expected for cluster-scale halos within
the CDM clustering scenario. We use the probability distribution
functions (PDFs) of the axis ratios, which are derived in JS02 using the
$N$-body simulations.  The PDFs are given as a function of redshift and
halo mass; a cluster-scale halo at $z\approx 0$ is typically
characterized by a triaxial shape with the mean value $a/c\approx 0.45$
and the standard deviation $\sigma(a/c)\approx 0.1$. For simplicity, we
add the following Gaussian prior to the $\chi^2$ (eq. [\ref{eqn:chi2}]): 
\begin{equation}
 \chi^2_{\rm ar}=-2.0\ln\left[p(a/c)p(a/b|a/c)\right],
\label{eqn:p_ac}
\end{equation}
where the subscript `ar' stands for axis ratios and the PDFs $p(a/c)$ and
$p(a/b|a/c)$ are given by equations (17) and (19) in JS02, respectively.
The resulting constraints are shown in Table \ref{table:fit} and Figures
\ref{fig:chi_tri} and \ref{fig:chi_mc}. However, the constraints are
only weakly tightened  and the main results we have so far found are not
largely changed: a broad range of the halo concentration or the virial
mass is still allowed.  

\begin{figure}
\epsscale{1.1} \plotone{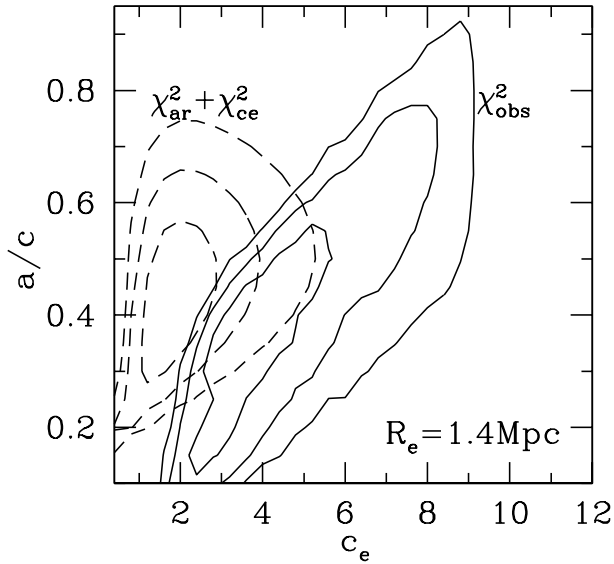} \caption{Constraint contours (solid
 lines) in the $c_e-a/c$ plane as in the previous figure, but the radius
 $R_e$ is fixed to $R_e=1.4{\rm Mpc}$ which is consistent  with
 observational constraints at $1-\sigma$ level (see Table
 \ref{table:fit}). Note that these results include the uncertainties of
 the other parameters besides $R_e$. The constraints are directly
 compared with the theoretical predictions (dashed lines) giving the
 $68\%$, $95\%$ and $99.7\%$ probabilities for the $c_e$ and $a/c$
 distribution derived using the model of JS02. While the PDF of $c_e$
 slightly depends on halo mass, in computing the theoretical predictions 
 we fix the virial mass to $M_{\rm vir}=1.5\times 10^{15}M_\odot$
 which roughly corresponds to  $R_e=1.4{\rm Mpc}$ for a given range of
 $(c_e,a/c,b/c)$  we have considered here. It is clear that the
 observation and theory are consistent at $\sim 2$-$\sigma$ level (see
 text for more details). 
 \label{fig:chi_tri_ce}}
\end{figure}
\begin{figure}
\epsscale{1.1} \plotone{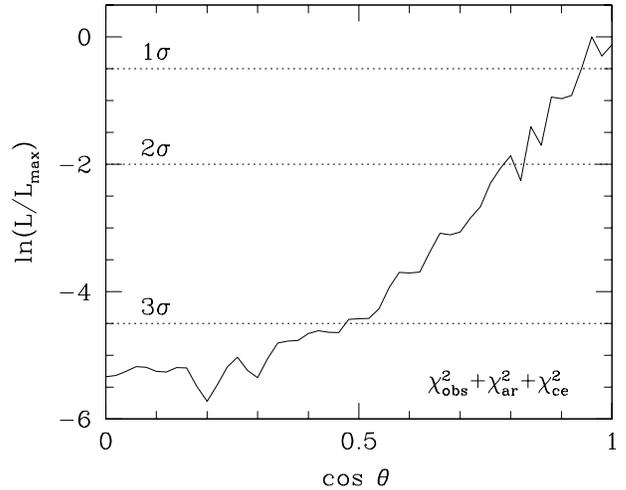} \caption{The likelihood function against
 the angle parameter $\theta$ between the major axis of triaxial halo
 and the line-of-sight direction, assuming the priors for the PDFs of
 axis ratios and halo concentration. The horizontal dotted lines show
 $68\%$, $95\%$, $99.7\%$ confidence limits ($\Delta\chi^2=1$, $4$ and
 $9$, respectively).  It is clear that an alignment of the halo
 elongation along the line-of-sight is favored in order for the
 CDM-based halo model to match the A1689 lensing measurements. 
 \label{fig:chi_theta}} 
\end{figure}
Hence, to more highlight the constraint on the halo concentration, 
Figure \ref{fig:chi_tri_ce} shows the error contours in the $c_e-a/c$
plane, when the radius $R_e$ is fixed to  $R_e=1.4{\rm Mpc}$ which is
consistent with observational constraints at 1-$\sigma$ level. Note that
the other parameters are allowed to vary. The clear trend, a smaller
$c_e$ is favored with decreasing $a/c$, is apparent. This result allows
a more direct comparison with the theoretical prediction, since JS02
suggested that the halo concentration $c_e$ be tightly correlated with
the axis ratio $a/c$. The PDF of $c_e$, $p(c_e)$, is given by equation
(20) in JS02 \citep[also see][]{oguri03}. The PDF is approximated with a
log-normal distribution, where the mean value of $c_e$ depends on halo
mass, $a/c$ and redshift (see eq. [23] in JS02) and the standard
deviation $\sigma(\ln c_e)\approx 0.3$. Combining with the PDFs
(eq. [\ref{eqn:p_ac}]) for the axis ratios, the dashed contours in Figure
\ref{fig:chi_tri_ce} show the 1-, 2- and 3-$\sigma$ probabilities for the
$c_e$ and $a/c$ parameters, expected for a halo at $z=0.18$, which are
estimated from $\chi^2_{\rm ar}$ + $\chi^2_{\rm ce} = 2.3$, $6.17$, and
$11.8$, respectively, where $\chi^2_{\rm ce}=-2\ln \left[p(c_e)\right]$
(the subscript `ce' stands for  the concentration parameter $c_e$). 
It is clear that the observation and theory overlap at $\sim 2$-$\sigma$
level. To see this more quantitatively, we compute the fraction of halos
whose axis ratio $a/c$ and concentration parameter $c_e$ fall inside the
contours constrained from the observation. We find that, if we take
$1$-, $2$-, and $3$-$\sigma$ limits as the observational constraints,
then the fraction becomes $2.7\%$, $6.1\%$, and $11.6\%$,
respectively. In contrast, if we {\em a priori} assume the spherical
halo model (B05a), the fraction is significantly reduced to $<0.1\%$,
even if we take the 3-$\sigma$ limit as observational
constraints. Hence, we conclude that the A1689 lensing measurements are
indeed  compatible with the CDM-based triaxial halo model, if the A1689
represents a rare population ($\approx 6\%$ fraction) of cluster-scale
halos (also see the discussion in the next section). 

Finally, we look at the constraints on the angle parameter $\theta$
between the major axis and the line-of-sight direction.  Figure
\ref{fig:chi_theta} shows the likelihood function against $\theta$,
derived from the observational constraints $\chi^2_{\rm obs}$ plus 
the theoretical PDFs of axis ratios and concentration parameter,
$\chi^2_{\rm ar}+\chi^2_{\rm ce}$. One can find $\cos\theta>0.8$
($\theta<37^\circ$) is obtained at 2-$\sigma$ level. Thus the major axis
of the triaxial halo, when the halo has 
 a reasonable concentration of
$c_e=1-2$, is favored to be aligned with the line-of-sight direction in
order to match the A1689 steep mass profile. 

\section{Summary and Discussion}
In this paper, we have employed the triaxial halo model of JS02 to
extract more reliable information on the three-dimensional mass
distribution of the massive cluster A1689, from the precise lensing
measurements done based on the ACS and Subaru data.  We have shown
several important results.  First, based on the simple thought
experiment in \S \ref{sec:test}, it was demonstrated that the halo
triaxiality  could cause a significant bias in estimating the virial
mass and concentration parameter from the lensing information, if a
spherical halo model is {\em a priori} assumed for the model fitting
(see Figure \ref{fig:chi}).  In particular, both the virial mass and
concentration parameter are  overestimated (underestimated) when the
triaxial halo is projected along the major (minor) axis. Second, we
derived constraints on the parameters of the triaxial halo model from
the fitting to the lensing measurements of A1689, by changing all 7
parameters that specify the triaxial model; the {\it virial} radius and
concentration parameter of the triaxial modeling, two axis ratio
parameters, and three angle parameters that determine the halo
orientation. We have shown that the halo triaxiality significantly
weakens the parameter constraints, compared to those derived when {\em a
priori} assuming the spherical halo model. In addition, the
parameter constraints are strongly degenerate, mainly because of a lack
of the halo shape information along the line-of-sight (see Table
\ref{table:fit} and Figures \ref{fig:chi_tri} and \ref{fig:chi_mc}). As
a result, the triaxial halo model with a theoretically expected halo
concentration $c_e=1-2$ is acceptable at 1-$\sigma$ level, while the halo
shape parameters (axis ratios and orientation) are hardly
constrained. To further improve the model constraints, we used the
theoretical predictions of probability distributions for the halo
shapes, derived in JS02, as the priors to the model fitting (see Figures
\ref{fig:chi_tri}, \ref{fig:chi_mc}, \ref{fig:chi_tri_ce} and
\ref{fig:chi_theta}). Most interestingly, we found that about $6\%$
population of cluster-scale halos, expected from the CDM structure
formation, can match the A1689 mass profile, if the major axis is almost
aligned with the line-of-sight direction. Hence, although it has been
shown that the triaxiality has a great impact on strong lensing
probabilities \citep{oguri03,dalal04,oguri04b}, our results clearly
demonstrate that the halo triaxiality is also very important for
exploring the three-dimensional mass distribution of an individual
cluster from the lensing measurements.  

It is encouraging that the triaxial halo model yields the
lensing-constrained mass of A1689 consistent with the mass estimate
derived from the $X$-ray observation at 3-$\sigma$ level (see Figure
\ref{fig:chi_mc}), while the significant discrepancy of factor 2 has
been reported in the literature (\citealt{miralda95,andersson04}; B05b).
We have checked that the consistent mass estimate appears when the major
axis of the triaxial halo is oriented along the line-of-sight direction,
as inferred from Figure \ref{fig:chi}.  Note that the $X$-ray mass
estimate is likely to be  less affected by the halo triaxiality than the
lensing, as described  in \citet{gavazzi05}. Therefore, it is
interesting to notice that the triaxial halo elongated along the
line-of-sight could resolve both the previously reported discrepancies
of the high concentration as well as of the lensing and $X$-ray mass
estimates. Unfortunately, however, this hypothesis cannot be further
tested by the lensing information alone we have used, because the halo
shape parameters (axis ratios and orientation) could not be well
constrained. Improving the halo shape constraints will allow us to make
a more severe, quantitative test of the triaxial halo model, as it is
clear from Figure \ref{fig:chi_tri}. This would be possible by combining
the lensing measurements with the detailed observations of kinematics,
$X$-ray emission and/or the Sunyaev-Zel'dovich (SZ) effect, since
projections affect the observables in different ways
\citep[e.g.,][]{girardi97,zaroubi01,marshall03}. For instance, as noted
above, the mass profile reconstructed from the $X$-ray surface
brightness profile is much less sensitive to the halo orientation, thus
by comparing two-dimensional profiles from $X$-ray and lensing we will
be able to obtain information on the degree of elongation along
line-of-sight. The SZ effect is also quite useful since its projection
effect is complementary to that of $X$-ray \citep{lee04}.

While our triaxial modeling is much more realistic than the simple
spherical fit, there is still room for improvement. A caveat is that the
PDFs of axis ratios and concentration (JS02) were derived for less
massive clusters ($M_{\rm vir}\lesssim 10^{14}M_\odot$). Thus it is
unclear if the extrapolated predictions still hold for very massive
clusters ($M_{\rm vir}\gtrsim 10^{15}M_\odot$) considered here. More
importantly, the selection effect, which we have neglected in this
paper, might be important. A1689 is well known as the strong-lensing
cluster, with the largest Einstein radius ($\sim 50''$). Strong lensing
probabilities are very sensitive to shapes and orientations of dark halos 
such that a halo with large triaxiality, high concentration and the
major axis aligned with line-of-sight has higher lensing probabilities
\citep{oguri03,dalal04,oguri04b}. Therefore A1689 might be indeed such a
cluster, as implied in this paper.  Thus taking into account the
selection effect could further reduce the discrepancy of A1689 
\citep[e.g.,][]{hennawi05}. However,
in practice it is quite difficult to quantify the selection effect.
 For strong lens systems discovered in a well-defined homogeneous
survey, such as the large-separation lensed quasar SDSS J1004+4112
\citep{inada03,oguri04a}, it will be possible to take this selection
effect into account in a correct manner. 

Another important implication we have found is that the halo triaxiality
can be a source of systematic errors in estimating the virial mass from
the lensing measurement. While it has been shown that the lensing-based
mass determination suffers from the projection effect due to an
intervening matter, not necessarily associated with a cluster of
interest \citep{metzler01,wanbsganss05}, our results imply that the halo
shape itself causes a bias in the mass determination \citep[also
see][]{bartelmann95,clowe04,hamana04}.  The accurate mass determination
is crucial to achieve the full potential of future cluster cosmology
experiments such as expected from Planck, where the cluster redshift
distribution is measured to constrain cosmological models
\citep{haiman01}. Therefore, it is again very interesting to address how
the mass determination can be improved for an individual cluster as well
as in a statistical sense by combining the lensing, $X$-ray and SZ
effect measurements.  

\acknowledgments 
We thank Jounghun Lee and Yipeng Jing 
for useful discussion. M.O. is supported by JSPS
through JSPS Research Fellowship for Young Scientists.   
This work has been supported in part by a Grand-in-Aid for Scientific
Research (17740129) of the Ministry of Education, Culture, Sports,
Science and Technology in Japan.


\end{document}